\documentclass[letterpaper]{article} 
\usepackage{aaai25}  
\usepackage{times}  
\usepackage{helvet}  
\usepackage{courier}  
\usepackage[hyphens]{url}  
\usepackage{graphicx} 
\usepackage{natbib}  
\usepackage{caption} 
\usepackage{algorithm}
\usepackage{algorithmic}

\usepackage{microtype}
\usepackage{float}
\usepackage{graphicx}
\usepackage{multirow}
\usepackage{amssymb}
\usepackage{amsmath}
\usepackage{newfloat}
\usepackage{listings}
\DeclareCaptionStyle{ruled}{labelfont=normalfont,labelsep=colon,strut=off} 
\floatstyle{ruled}
\newfloat{listing}{tb}{lst}{}
\floatname{listing}{Listing}
\title{Mental-Perceiver: Audio-Textual Multi-Modal Learning for Estimating Mental Disorders}
\author{
Jinghui Qin\textsuperscript{\rm 1}, Changsong Liu\textsuperscript{\rm 2,3}, Tianchi Tang\textsuperscript{\rm 2}, Dahuang Liu\textsuperscript{\rm 2},  \\ Minghao Wang\textsuperscript{\rm 2}, Qianying Huang\textsuperscript{\rm 2}, Rumin Zhang\textsuperscript{\rm 2,4}\thanks{Corresponding author}
}
\affiliations{
    \textsuperscript{\rm 1}Guangdong University of Technology\\
    \textsuperscript{\rm 2}Guangdong Shuye Intelligent Technology Co., Ltd.\\
    \textsuperscript{\rm 3}University of Toronto\\
    \textsuperscript{\rm 4}Ningbo Institute of Digital Twin, Eastern Institute of Technology, Ningbo, China

    \{scape1989, liucs.oise\}@gmail.com, \{tangtc, liudh, wangmh, huangqy\}@gdshuyeit.com, rzhang01@idt.eitech.edu.cn

}

\usepackage{bibentry}

\begin{document}

\maketitle

\begin{abstract}
Mental disorders, such as anxiety and depression, have become a global concern that affects people of all ages. Early detection and treatment are crucial to mitigate the negative effects these disorders can have on daily life. Although AI-based detection methods show promise, progress is hindered by the lack of publicly available large-scale datasets. To address this, we introduce the \textbf{M}ulti-\textbf{M}odal \textbf{Psy}chological assessment corpus (MMPsy), 
a large-scale dataset containing audio recordings and transcripts from Mandarin-speaking adolescents undergoing automated anxiety/depression assessment interviews. MMPsy also includes self-reported anxiety/depression evaluations using standardized psychological questionnaires. Leveraging this dataset, we propose Mental-Perceiver, a deep learning model for estimating mental disorders from audio and textual data. Extensive experiments on MMPsy and the DAIC-WOZ dataset demonstrate the effectiveness of Mental-Perceiver in anxiety and depression detection. 
\end{abstract}

%
\begin{links}
    \link{Datasets}{https://github.com/shuyeit/mmpsy-data}
\end{links}

\section{Introduction}

Anxiety and depression are prevalent mental disorders that can have a significant and detrimental impact on an individual's life. If left untreated, these disorders can disrupt work, study, and social interactions. Symptoms include persistent negative emotions, behavioral changes, and physiological disturbances~\cite{cowen2012shorter}. However, misdiagnosis is common due to the overlap of symptoms with other physical and physiological conditions. The World Health Organization (WHO) 2019 report~\cite{de2019world} indicates that more than 301 million people, including 58 million children and adolescents, suffer from anxiety disorders. In addition, more than 280 million people, including 23 million children and adolescents, are affected by depression. Despite the prevalence of these conditions, the treatment rate for both anxiety and depression remains notably low due to the substantial time and financial commitments required for diagnosis and treatment~\cite{kessler2012costs}. Furthermore, individuals, particularly children and adolescents, can conceal their true mental state during assessments and interviews, preventing an accurate diagnosis.

To address this challenging issue, automatic detection of anxiety and depression offers a potential solution. The development of a suitable psychological assessment corpus related to anxiety and depression and the construction of an automated anxiety and depression detection system are essential. This will allow individuals to assess their anxious or depressive states privately and increase their willingness to consult with psychologists after conducting self-assessments. Furthermore, such a psychological assessment corpus and automated detection system would greatly assist psychologists in diagnosing anxiety and depression disorders, even when patients conceal their true mental states.

A significant challenge facing current research in automatic anxiety and depression detection is the absence of extensive datasets to train and evaluate machine learning models. Existing methodologies frequently depend on the judgment of expert psychologists (e.g., during clinical interviews) for data collection, making the creation of large-scale annotated datasets costly. Publicly accessible datasets, such as DAIC-WOZ~\cite{gratch2014distress} and AViD-Corpus~\cite{valstar2014avec}, are restricted in scale and scope. Although initiatives like EATD-Corpus~\cite{shen2022automatic} have been undertaken using self-reported mental health questionnaires, challenges persist regarding data size and validation.

In this work, our goal is to address the challenges in facilitating automatic detection of anxiety and depression and to extend the research boundaries to a more real world setting. First, we introduce a large-scale \textbf{M}ulti-\textbf{M}odal \textbf{Psy}chological assessment corpus (MMPsy) about anxiety and depression in Mandarin-speaking adolescents. The MMPsy comprises audio and extracted transcripts of responses from anxious/depressed and non-anxious/non-depressed adolescent volunteers. 
After data pre-processing and cleaning, 7,736 cleaned interview data are available for anxiety detection and 4,247 for depression detection.
To our knowledge, MMPsy is the first publicly available adolescent psychological assessment corpus capable of simultaneously detecting anxiety and depression, and containing both audio and text data in Chinese.

Subsequently, we propose a deep learning model, termed \textbf{Mental-Perceiver}, to automatically detect anxious/depressive mental states based on users' audio and corresponding transcripts. Mental-Perceiver first maps multi-modal inputs and category semantic priors to a fixed-size feature space with learnable embeddings for multi-modal fusion. It then employs a fully attentional network~\cite{jaegle2021perceiver} to further process the fused multimodal features and decode the underlying mental state with a learnable query array. Extensive experiments on our MMPsy and DAIC-WOZ datasets demonstrate the effectiveness and superiority of our proposed Mental-Perceiver model for the estimation of anxiety and depression. 
The key contributions of this work are summarized as follows:
\begin{itemize}
\item We introduce MMPsy, a large-scale multi-modal corpus for mental health assessment.
\item We propose Mental-Perceiver, a fully attentional network for automated anxiety/depression detection from audio and transcripts.
\item Extensive experiments on the MMPsy and DAIC-WOZ datasets demonstrate the effectiveness of the Mental-Perceiver.
\end{itemize}


\section{Related Work}

\subsection{Anxiety/Depression Detection Datasets}
Publicly available datasets for the detection of anxiety and depression remain limited due to the challenges of collecting sensitive mental health data. Existing datasets, such as DAIC-WOZ~\cite{gratch2014distress} and AViD-Corpus~\cite{valstar2014avec}, focus primarily on depression and often lack textual transcripts or sufficient sample sizes. The MODMA dataset~\cite{Cai_2022}, while incorporating EEG signals along with audio and video recordings, presents challenges for real-world application due to the difficulty in collecting EEG data in regular scenarios and its limited sample size. Recent efforts like EATD-Corpus~\cite{shen2022automatic} have attempted to address these limitations by incorporating self-reported measures and textual data, but are constrained by a relatively small sample size. This scarcity of diverse and large-scale datasets hinders the development and evaluation of robust machine learning models for anxiety and depression detection, particularly in multilingual and cross-cultural contexts.

\subsection{Automatic Anxiety/Depression Detection}

Early research in automatic anxiety/depression detection focused on extracting relevant features from interview responses. Williamson et al.~\cite{williamson2016detecting} utilized semantic cues from voice, facial action units, and transcribed text within a Gaussian Staircase Model for depression detection. Yang et al.~\cite{yang2016decision} employed manual transcript analysis to select depression-related questions and built a decision tree for prediction. Similarly, Sun et al.~\cite{sun2017random} used content analysis to select text features from interview transcripts and applied Random Forest to detect depression tendency. Gong et al.~\cite{gong2017topic} utilized topic modeling for feature selection, while Giannakakis et al.~\cite{giannakakis2017stress} investigated the correlations between facial cues and perceived stress/anxiety.

Deep learning advancements have enabled more effective multi-modal feature extraction and integration. Ma et al.~\cite{ma2016depaudionet} used CNN-LSTM models for depressive audio encoding. Yang et al.~\cite{yang2017multimodal} used deep CNNs with audiovisual descriptors for the detection of depression. Tuka et al.~\cite{al2018detecting} leveraged Pearson coefficients to select relevant audio and text characteristics for the evaluation of depression based on LSTM. Haque et al.~\cite{haque2018measuring} proposed a causal CNN to generate embeddings from acoustic, visual, and linguistic features for the prediction of depression. Shen et al.~\cite{shen2022automatic} and Lin et al.~\cite{lin2022deep} focused on speech and linguistic content for depression detection, while Agarwal~\cite{agarwal2023detecting} explored machine learning for anxiety diagnosis from audio journals.

Recent years have witnessed significant advances in transformer-based and LLM approaches for the detection of mental disorders. Ji et al.~\cite{ji2021mentalbert} introduced domain-specific models, MentalBERT and MentalRoBERTa, which demonstrated superior performance on mental health benchmarks compared to general-purpose LLMs. Xu et al.~\cite{xu2023instruction} revealed that instruction-tuned smaller LLMs could achieve higher balanced accuracy than larger models in mental health prediction tasks, while Binz and Schulz~\cite{binz2023turning} enhanced LLMs' emotional state modeling capabilities by fine-tuning with psychological experiment data. In parallel, Qi et al.~\cite{qi2023supervised} achieved notable improvements in cognitive distortion and suicide risk detection using fine-tuned GPT-3.5 models. The field has increasingly embraced multi-modal approaches, as evidenced by Ahmed et al.~\cite{ahmed2023taking}, who developed a comprehensive framework that incorporates video, audio, text and EEG data with uncertainty management capabilities. Building on this trend, Ding et al.~\cite{ding2024intervoxnet} proposed IntervoxNet, a dual-modal audio-text fusion model using transformer architectures, achieving an impressive F1 score of 0.90 in the detection of depression from interview data. 

\section{MMPsy: A New Benchmark}

\begin{table}[t]
\centering
\resizebox{0.8\linewidth}{!}{
\begin{tabular}{lcc}
\hline
\textbf{Characteristic} & \textbf{Anxiety Subset} & \textbf{Depression Subset} \\
\hline
\multicolumn{3}{l}{\textbf{Gender}} \\
Males & 2,262 (53.02\%) & 4,087 (52.68\%) \\
Females & 2,004 (46.98\%) & 3,671 (47.32\%) \\
\hline
\multicolumn{3}{l}{\textbf{Grade Level}} \\
Grade 4 & 335 & - \\
Grade 5 & 987 & - \\
Grade 6 & 1,342 & - \\
Grade 7 & 1,646 & 1,564 \\
Grade 8 & 1,554 & 1,397 \\
Grade 9 & 1,579 & 1,305 \\
Grades 10-12 & 315 & - \\
\hline
\end{tabular}
}
\caption{Demographic Information of Data Collection Participants}
\label{tab:demographics}
\end{table}

Data collection was conducted during the mandated annual mental health assessment of elementary and middle school students in Guangdong Province, China, as required by local government regulations\footnote{Guangzhou Municipal Education Bureau Website - Regulations on Promoting Mental Health of Primary and Secondary School Students in Guangzhou (Full Text): \url{https://jyj.gz.gov.cn/yw/jyyw/content/post_9907147.html}}. The research team received approval from schools and their administrative Education Bureaus to provide mental health assessment services and collect data for research purposes. Under institutional supervision, data were collected from over 10,000 primary and secondary school students. All participants provided informed consent with guidance from supervising teachers or social workers. Demographic characteristics of the participants are presented in Table~\ref{tab:demographics}.

Participants were asked to provide spontaneous responses to 10 specifically designed questions addressing anxiety and depression. Additionally, they completed either the GAD-7~\cite{mossman2017generalized} or PHQ-9~\cite{kroenke2001phq} questionnaire for respective disorder detection. The GAD-7, comprising seven items measuring anxiety severity, is a standardized screening tool widely used in clinical practice. The PHQ-9, consisting of nine items measuring depression severity, serves a similar function for depression screening. For Chinese populations, scores $\geq$ 10 on either scale indicate the presence of the respective disorder. Based on these criteria, the anxiety subset of MMPsy includes 704 anxious and 7,032 non-anxious participants, while the depression subset comprises 853 depressed and 3,394 non-depressed participants.

\begin{table}[t]
\centering
\resizebox{0.9\linewidth}{!}{
\begin{tabular}{c|c|c|c|c}
\hline
Parts &Subset & $\times$ & $\checkmark$ & Avg Duration (Sec)\\ \hline
\multirow{2}{*}{Anxiety} &Train &5625 & 563  &68.05\\
                         &Validation &704 & 70 &65.97\\
                         &Test &703 &71 &68.67\\\hline
                         
\multirow{2}{*}{Depression} &Train &2715 &682 &59.05\\
                            &Validation  &340  &85 &55.27\\
                            &Test &339 &86 &61.76\\\hline
\end{tabular}
}
\caption{Statistics for our MMPsy dataset. $\times$ means the label is non-anxious or non-depressed and $\checkmark$ denotes the data label is anxious or depressed.}
\label{table1}
\end{table}

The construction of MMPsy comprised two primary phases: data collection and preprocessing.
\begin{itemize}
    \item \textbf{Data collection}. A custom web application was developed to conduct interviews and collect audio responses and questionnaire data. The application administered 10 questions along with the GAD-7 or PHQ-9 questionnaires. The audio responses were automatically recorded and uploaded to the server along with the questionnaire results. This phase yielded 17,247 raw interviews for anxiety and 11,306 for depression. To ensure the validity of the responses, additional control questions were embedded within the questionnaires to assess the reliability of the responses.
    
    \item \textbf{Data preprocessing}. The raw interview data underwent several preprocessing steps. Initially, responses were filtered based on the control questions to eliminate inauthentic entries. Subsequently, mute recordings and those shorter than 1 second were removed, and silent segments were eliminated using voice activity detection. Background noise was then removed using Spleeter~\cite{hennequin2020spleeter}. Textual transcripts were extracted using Paraformer~\cite{gao2022paraformer} and manually verified through audio comparison to ensure semantic consistency at the word level.

    \item \textbf{Data anonymization and reporting}. To protect participant privacy, all sensitive information and any data that could potentially identify participants were meticulously removed during the preprocessing stage. Anonymization was carried out in compliance with ethical guidelines and data protection regulations. Additionally, for students whose responses on the GAD-7 or PHQ-9 scales indicated potential anxiety or depression problems, their data was securely reported to the designated school administrative offices, as required by institutional and regulatory policies, ensuring appropriate follow-up support.
\end{itemize}

The preprocessing phase yielded 7,736 cleaned interviews for anxiety detection and 4,247 for depression detection. These data were randomly partitioned into training, validation, and test sets using an 8:1:1 ratio. 
Consequently, the anxiety detection subset comprises 6,188 training, 774 validation, and 774 test participants, while the depression detection subset contains 3,397 training, 425 validation, and 425 test participants. 
The complete data statistics are presented in Table~\ref{table1}. Following the methodology of~\cite{wei2022multi}, each participant's audio and transcript data were sequentially organized and segmented into 60-second intervals with 10-second overlaps, generating multiple samples per participant.


\begin{figure*}[t]
\centering
\includegraphics[width=\textwidth]{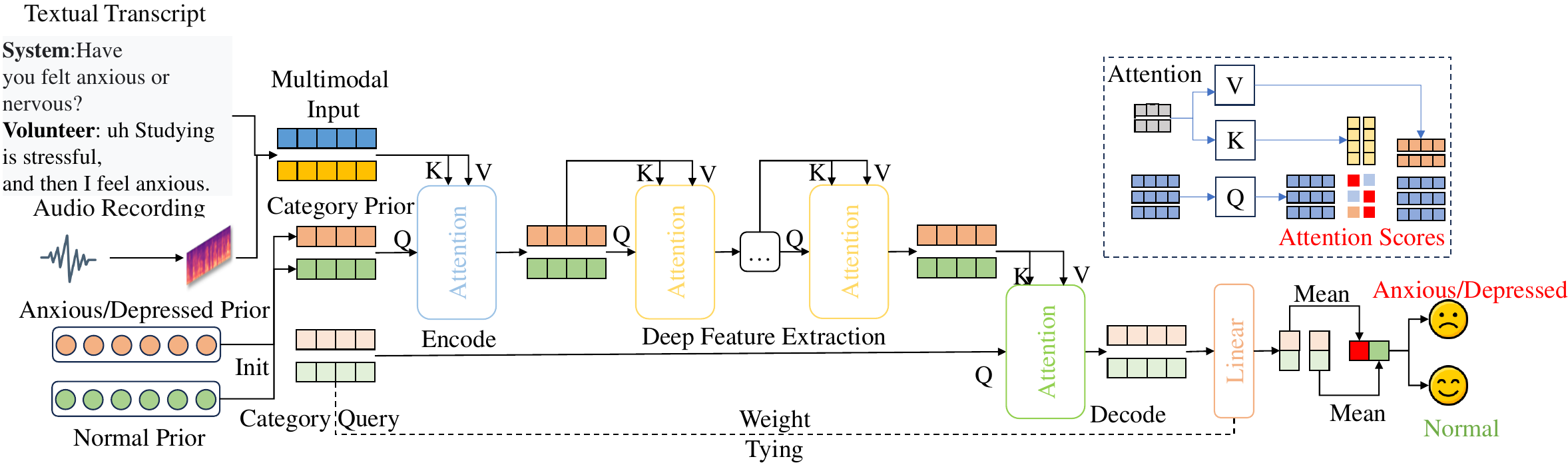}
\caption{Illustration of Mental-Perceiver.}
\label{fig_1}
\end{figure*}

\section{Mental-Perceiver}
This section describes the architecture of our proposed Mental-Perceiver as shown in Figure~\ref{fig_1}. We first encode by applying an attention module that maps multimodal input $x \in \mathbb{R}^{M \times D_x}$ to features in a latent space $z \in \mathbb{R}^{2 \times D_z}$ by interacting with a category prior $p \in \mathbb{R}^{2 \times D_p}$ which is obtained by computing the center point of different text representations from different categories. Then, we perform deep feature extraction on latent features $z$ by applying a series of attention modules that take in and return features $z'$ in this latent feature space. Finally, we decode by applying an attention module that maps latent arrays $z'$ and the query array $q \in \mathbb{R}^{2 \times D_q} $to the final feature representation $y \in \mathbb{R}^{2 \times D_y}$. Based on the final feature representation $y$, we apply a linear layer to map the feature $y$ to class-wise logit outputs $y^{C_0}\in \mathbb{R}^{1 \times 2}$ and $y^{C_1}\in \mathbb{R}^{1 \times 2}$. $M$ is the length of multimodal input, while $D_x$, $D_z$, $D_q$, and $D_y$ denote the feature dimension. With class-wise logits $y^{C_0}$ and $y^{C_1}$, to predict the class $c'$ of the input $x$, we compute the mean values of corresponding vector elements in the $y^{C_0}$ and $y^{C_1}$ to obtain final logits $y'$ followed by a Softmax function.

\subsection{Basic Attention module}
Following the pioneering work PerceiverIO~\cite{jaegle2021perceiver}, all attention modules deployed in our Mental-Perceiver are implemented as Transformer-style attention~\cite{attenallyouneed}. Each attention module applies a global query-key-value (QKV) attention operation followed by a multi-layer perceptron (MLP). The MLP is applied independently to each element of the index dimension. Both the encoder and decoder take in two input feature arrays. The first array is used as input to the attention module's key and value networks, and another array is used as input to the module's query network to interact and fuse with the first array. The output of the attention module has the same index dimension (the same number of elements) as the query input. Therefore, the attention module can be modeled as follows:
\begin{equation}
\text{attention}(Q, K, V) = MLP( Softmax \left( \frac{QK^T}{\sqrt{d_k}} \right) V)
\end{equation}

\subsection{Category Semantic Prior}
Our Mental-Perceiver extracts deep features based on the latent features $z$ which is produced by fusing the multimodal input $x$ and the latent embedding $p$ with an attention module. $p$ is used as a query and $x$ is used as the key and value. This means that the output $z$ is the result that $p$ extracts and fuse semantic information from $x$. So, the initialization of the latent embedding $p$ is crucial for learning more discriminative features in the following steps to identify a user's mental state according to the user's audio and transcript text. Semantic prior has been shown to help greatly learn more discriminative features~\cite{teney2019incorporating, dai2023pfemed, ding2023exploring}. Therefore, we build semantic priors for different categories, one for the normal category and another for the mental disorder category. Then, we use these two semantic priors to initialize the latent embedding $p \in \mathbb{R}^{2 \times D}$ with a learnable MLP for mapping the hidden size of semantic priors to $D$. In Mental-Perceiver, we fixed the parameters of these two semantic priors and only optimized the parameters of the learnable MLP semantic priors. To obtain the semantic priors, we simply compute the center points for different categories.

Formally, given the text representation set $\textbf{E}_t \in \mathbb{R}^{N \times H}$ of a category on the training set, where $N$ is the number of data samples for the current category and  $H$ is the hidden size of a text representation, we can compute the semantic prior $p^{C_i} \in \mathbb{R}^{1 \times H} $ by averaging the normalized $\textbf{E}_t$ at the index dimension. This procedure can be modeled as follows:
\begin{equation}
p^{C_i} = Avg(Norm(\textbf{E}_t))
\label{p_eq}
\end{equation}
where $Avg$ is the averaging function and $Norm$ is the z-score normalization function that normalizes each vector separately in $\textbf{E}_t$. Let $C_0$ and $C_1$ denote the classes of normal people and psychological patients, respectively, we first obtain two semantic priors $p^{C_0}$ and $p^{C_1}$ according to Equation (\ref{p_eq}). Then, we obtain $p$ by concatenating $p^{C_0}$ and $p^{C_1}$ at the index dimension followed by a learnable MLP layer as follows:
\begin{equation}
p = MLP(p^{C_0} \odot p^{C_1})
\end{equation}
where $\odot$ denotes the concatenation.

\subsection{Encoder}
The encoder, consisting of an attention module, takes charge of mapping the multimodal input $x$ into latent features $z$ by utilizing category semantic prior-enhanced latent embedding $p$ to query the multimodal input $x$. This way will fuse the multimodal input $x$ and category semantic prior-enhanced latent embedding $p$ and build fused prior-guided latent features $z$, which will be used as input to the next deep feature extraction. The encoder can be modeled as follows:
\begin{equation}
z = attention(p, x, x)
\end{equation}

\subsection{Deep Feature Extraction}
Once we obtain the latent features $z$, we conduct deep feature extraction based on the input latents $z$ by applying a series of attention modules that take in and return latents $z_i$ in this latent space iteratively.
This module can be modeled as follows:
\begin{equation}
\begin{aligned}
z_1 &= attention(z, z, z) \\
z_2 &= attention(z_1, z_1, z_1) \\
&... \\ 
z_k &= attention(z_{k-1}, z_{k-1}, z_{k-1}) \\
\end{aligned}
\end{equation}
where $k$ is a hyper-parameter that is simply set to 8.

\subsection{Decoder}
The goal of the decoder is to produce a final class-wise logit output of size $2 \times 2$, given a latent representation of size $2 \times D_z$. Let $z' = z_k$, the decoder first applies an attention module to map the latent $z'$ to output features $y$. Then, the decoder applies a linear layer to map $y$ into the final class-wise logit output $[y^{C_0}, y^{C_1}] \in \mathbb{R}^{2\times2}$, where $y^{C_0} \in \mathbb{R}^{1\times2}$ and $y^{C_1} \in \mathbb{R}^{1\times2}$ are two logit outputs for indicating whether the multimodal input $x$ matches with semantic prior $p^{C_0}$ and $p^{C_1}$. Finally, we compute the mean vector of these two output logit vectors $y^{C_0}$ and $ y^{C_1}$ at the index dimension as the final classification logits $y'$. Therefore, the decoder can be modeled as follows:
\begin{equation}
\begin{aligned}
y &= attention(q, z', z') \\
[y^{C_0}, y^{C_1}] &= Linear(y) \\
y' &= Mean(y^{C_0} \odot y^{C_1})
\end{aligned}
\end{equation}

\subsection{Training Objectives}
To optimize our Mental-Perceiver, we deploy two losses. The first one is the matching loss $\mathcal{L}_{match}$ and another is the classification loss $\mathcal{L}_{cls}$. Both these two losses are binary cross-entropy loss functions. The matching loss aims to optimize the matching degree between the multimodal input $x$ and its corresponding category semantic prior while the classification loss optimizes the model to be able to identify the inherent mental state according to multimodal input $x$ with the help of category prior $p$ and category query $q$. 

Formally, given a  multimodal input $x$ and its class $C_x$ which can be 0 or 1, the training objectives can be modeled as follows:
\begin{equation}
\begin{aligned}
\mathcal{L}_{match} = & -C_x (log y^{C_0}_0+log y^{C_1}_1) \\ &- (1-C_x) (log y^{C_0}_1 + log y^{C_1}_0) \\
\mathcal{L}_{cls} = & - C_x logy'_0 - (1-C_x) logy'_1 \\
\mathcal{L} = &  \mathcal{L}_{match} + \mathcal{L}_{cls}
\end{aligned}
\end{equation}
where $y^{C_0}_0$ and $y^{C_0}_1$ are the 0-th element and 1-th element in the probability distribution obtained from class-wise logits $y^{C_0} \in \mathbb{R}^{1\times2}$ by Softmax function while $y^{C_1}_0$ and $y^{C_1}_1$ are the 0-th element and 1-th element in the probability distribution obtained from class-wise logits $y^{C_1} \in \mathbb{R}^{1\times2}$  by Softmax function. Similarly, $y'_0$ and $y'_1$ are the two probabilities on normal class (0) and mental disorder class (1) that can be obtained by applying the Softmax function on $y'$.   

\section{Experiments}

\subsection{Datasets}
We conduct experiments on both two subsets of MMPsy for anxiety detection and depression detection. We use MMPsy-Anxiety and MMPsy-Depression to represent these two subsets. Besides, we also verify our Mental-Perceiver on the Distress Analysis Interview Corpus - Wizard of Oz (DAIC-WOZ) dataset~\cite{gratch2014distress}. DAIC-WOZ contains clinical interviews of 189 participants designed to support the diagnosis of psychological distress conditions such as anxiety, depression, and posttraumatic stress disorder (PTSD). During each interview, several data in different formats as well as modalities are recorded simultaneously. However, only the acoustic recordings and transcriptions are chosen in this work for a fair comparison. Moreover, the given GT is an eight-item Patient Health Questionnaire depression scale (PHQ-8), which indicates the severity of depression. A PHQ-8 Score $\geq$ 10 implies that the participant is undergoing a mental disorder. 

\subsection{Baselines}
The main baselines to be compared are listed as follows:
\begin{itemize}
\item \textbf{SVM}~\cite{Pedregosa2011ScikitlearnML}: a robust shallow model capable of performing binary classification efficiently by using a kernel trick, 
which transforms original datapoints into coordinates in the higher dimensional feature space. 
\item \textbf{RandomForest}~\cite{Pedregosa2011ScikitlearnML}: it is a robust ensemble learning method for classification by constructing a multitude of decision trees at training time. For classification tasks, the output of the random forest is the class selected by most trees.  
\item \textbf{XGBoost}~\cite{Chen2016XGBoostAS}: it is a robust toolbox for classification via an optimized distributed gradient boosting. 
\item \textbf{NUSD}~\cite{Wang2023NonuniformSD}: it is a deep model ECAPA-TDNN enhanced with a speaker disentanglement method that utilizes a non-uniform mechanism of adversarial SID loss maximization.
\item \textbf{ConvLSTM}~\cite{wei2022multi}: it is a Convolutional Bidirectional LSTM with a sub-attention mechanism for linking heterogeneous information.
\item \textbf{PerceiverIO}~\cite{jaegle2021perceiver}: it is a general-purpose architecture that handles multimodal data with fully attention design and a flexible querying mechanism.
\item \textbf{AFABNet}~\cite{xu2024attention}: it is an attention-based acoustic feature fusion network for depression detection by combining four different acoustic features.
\item \textbf{Qwen2-Audio-Instruct}~\cite{chu2024qwen2}: it is a Qwen large audio-language model (7B) that can accept various audio signal inputs and perform audio analysis or direct textual responses with regard to speech instructions.
\end{itemize}
For the shallow models SVM, RandomForest, and XGBoost, we provide the following audio features as input: F0 statistics (mean), log-energy, zero-crossing-rate, loudness, pitch period entropy, jitters, shimmers, harmonics-to-noise ratio, detrended fluctuation analysis, linear spectral coefficients-0, linear spectral frequencies-0, formants (F1), and amplitude Shannon entropy. All these features can be extracted by applying Surfboard~\footnote{https://github.com/novoic/surfboard}~\cite{Lenain2020SurfboardAF}. Besides, we also extract topic words by TF-IDF as text features for these shallow models. For the deep models, we use BERT~\cite{devlin-etal-2019-bert} to extract features and use Mel-spectrum to represent audio. 




\begin{table*}[t]
\centering
\resizebox{0.95\textwidth}{!}{
\begin{tabular}{c|c|cccc|ccc|ccc}
\hline
Datasets & Methods &Acc & UAR & Sens &Spec  & Precision & Recall & F1 & Precision & Recall & F1 \\ \hline
\multicolumn{6}{c}{} & \multicolumn{3}{c}{Normal}  & \multicolumn{3}{c}{Anxious} \\ \hline
\multirow{6}{*}{MMPsy-Anxiety}  & SVM &0.83 & 0.64 & 0.41 &0.87  & 0.94 & 0.87 & 0.9 & 0.24 & 0.41 & 0.3 \\ 
& RandomForest &0.82 & 0.63 & 0.39 &0.87  & 0.93 & 0.87 & 0.9 & 0.23 & 0.39 & 0.29 \\
& XGBoost &0.88 & 0.6 & 0.25 &\textbf{0.94}  & 0.93 & \textbf{0.94} & \textbf{0.93} & 0.3 & 0.25 & 0.27 \\
& NUSD &0.79 & 0.51 & 0.17 &0.85  & 0.91 & 0.85 & 0.88 & 0.1 & 0.17 & 0.13 \\
& ConvLSTM &0.83 & 0.55 & 0.21 & 0.9  & 0.92 & 0.9 & 0.91 & 0.18 & 0.21 & 0.19 \\
& PerceiverIO &0.83 & 0.74 & 0.63 & 0.84  & 0.95 & 0.84 & 0.90 & 0.29 & 0.63 & 0.40 \\
& AFABNet &0.82	&0.63 &0.41 &0.86 &0.93 &0.86 &0.89 &0.22 &0.41 &0.29\\
& Qwen2-Audio-Instruct &0.72 &0.76 &\textbf{0.80} &0.71 &\textbf{0.97}  &0.71  &0.82  &0.22 &\textbf{0.80} &0.34\\
& Mental-Perceiver (Ours) & \textbf{0.85} & \textbf{0.76} & 0.65 &0.87  & 0.96 & 0.87 & 0.92 & \textbf{0.34} & 0.65 & \textbf{0.45} \\
\hline
\multicolumn{6}{c}{} & \multicolumn{3}{c}{Normal}  & \multicolumn{3}{c}{Depressed} \\ \hline
\multirow{6}{*}{MMPsy-Depression} & SVM &0.76 & 0.71 & 0.63 &0.79  & 0.89 & 0.79 & 0.84 & 0.43 & 0.63 & 0.51 \\ 
& RandomForest &0.76 & 0.63 & 0.41 & 0.85  & 0.85 & 0.85 & 0.85 & 0.41 & 0.41 & 0.41 \\
& XGBoost &0.78 & 0.64 & 0.41 &0.88  & 0.85 &  0.88 &  0.87 & 0.46 & 0.41 & 0.43 \\
& NUSD &0.67 & 0.51 & 0.26 &0.78  & 0.8 & 0.78 & 0.79 & 0.22 & 0.26 & 0.24 \\
& ConvLSTM &0.77 & 0.53 & 0.14 &0.93  & 0.81 & 0.93 & 0.86 & 0.32 & 0.14 & 0.2 \\
& PerceiverIO &0.81 & 0.77 & 0.66 & 0.85  & 0.91 & 0.85 & 0.88 & 0.53 & 0.66 & 0.59 \\
& AFABNet &0.78	&0.58 &0.24 &0.92 &0.83 &0.92 &0.87 &0.44 &0.24 &0.31 \\
& Qwen2-Audio-Instruct  &0.79 &0.58 &0.21 &\textbf{0.94} &0.82  &\textbf{0.94}  &0.88   &0.45  &0.21  &0.29\\
& Mental-Perceiver (Ours) &\textbf{0.85} & \textbf{0.79} & \textbf{0.69} &0.89  & \textbf{0.92} & 0.89 & \textbf{0.9} & \textbf{0.61} & \textbf{0.69} & \textbf{0.64} \\
\hline
\multicolumn{6}{c}{} & \multicolumn{3}{c}{Normal}  & \multicolumn{3}{c}{Depressed} \\ \hline
\multirow{6}{*}{DAIC-WOZ}  & SVM &0.38 & 0.35 & 0.29 &0.42  & 0.58 & 0.42 & 0.49 & 0.17 & 0.29 & 0.22 \\ 
& RandomForest &0.7 & 0.54 & 0.14 &0.94  & 0.72 & 0.94 & 0.82 & 0.5 & 0.14 & 0.22 \\
& XGBoost &0.6 & 0.53 & 0.36 &0.7  & 0.72 & 0.7 & 0.71 & 0.33 & 0.36 & 0.34 \\
& NUSD &0.55 & 0.46 &  0.21 &0.7  & 0.68 & 0.7 & 0.69 & 0.23 & 0.21 & 0.22 \\
& ConvLSTM &0.4 & 0.57 & \textbf{0.88} &0.22  & \textbf{0.83} & 0.22 & 0.35 & 0.3 & \textbf{0.88} & 0.45 \\
& PerceiverIO &0.7 & 0.58 & 0.29 & 0.88  & 0.74 & 0.88 & 0.81 & 0.5 & 0.29 & 0.36 \\
& AFABNet  &0.74 &0.63 &0.36 &0.91 &0.77 &0.91 &0.83 &0.62 &0.36 &0.45 \\
& Qwen2-Audio-Instruct &0.70 &0.65 &0.52 &0.78 &0.79 &0.78 &0.79 &0.50 &0.52 &\textbf{0.51}\\
& Mental-Perceiver (Ours) &\textbf{0.79} & \textbf{0.66} & 0.36 &\textbf{0.97}  & 0.78 & \textbf{0.97} & \textbf{0.86} & \textbf{0.83} & 0.36 & 0.5 \\
\hline

\end{tabular}
}
\caption{Performance Comparison with our Mental-Perceiver and various baselines on MMPsy and DAIC-WOZ. The best result is highlighted in \textbf{bold}.}
\label{mainres}
\end{table*}

\subsection{Metrics}
In classification tasks, True Positive (TP), True Negative (TN), False Positive (FP), and False Negative (FN) from the confusion matrix are metrics used to measure the accuracy of model predictions. TP refers to instances where the model correctly predicted them as the positive class, while TN refers to instances where the model correctly predicted them as the negative class. On the contrary, FP represents cases where the model incorrectly labeled instances as positive, and FN denotes instances of the positive class that were incorrectly classified as negative. Based on these concepts, several common performance metrics are adopted to assess the model’s performance:
\begin{itemize}
\item \textbf{Accuracy}: Accuracy (Acc) represents the proportion of all predictions that are correctly predicted.
\item \textbf{Recall}: Recall measures the model’s ability to identify
all true positive instances, that is, the proportion of actual positives that are correctly identified. 
\item \textbf{Precision}: Precision denotes the proportion of samples
predicted as positive by the model that are actually positive,
focusing on the accuracy of positive predictions.
\item \textbf{F1-Score}: F1 Score ranges from 1 to 0, with a value closer
to 1 indicating better model performance, particularly in scenarios where the distribution of positive and negative samples is imbalanced. The F1 Score serves as a more comprehensive evaluation metric under such conditions.
\item \textbf{Sensitivity}: Sensitivity (Sens), also termed
true positive rate, is the ratio of positive predictions to the number of actual positives.  It is hence identical to the recall of the positive class,
Recall(1). 
\item \textbf{Specificity}: Specificity (Spec) is the ratio of negative predictions to the number of actual negatives, and therefore identical to Recall(0).
\item \textbf{UAR}: Unlike accuracy, which is biased by data imbalance, Unweighted Average Recall (UAR) is preferred in fields with imbalanced datasets, such as biomedical and paralinguistics. UAR, i.e., the average of Sensitivity and Specificity, provides a more balanced performance measure.
\end{itemize}
During these metrics, we use Accuracy (Acc), Unweighted
Average Recall (UAR),  Sensitivity (Sens), and Specificity (Spec) as the main evaluation metrics for evaluating overall model performance. Meanwhile, we also report the Recall, Precision, and F1-score separately for different categories. 

\begin{table*}[t]
\centering
\resizebox{0.9\textwidth}{!}{
\begin{tabular}{c|c|cccc|ccc|ccc}
\hline
Datasets & Modalities &Acc & UAR & Sens &Spec  & Precision & Recall & F1 & Precision & Recall & F1 \\ \hline
\multicolumn{6}{c}{} & \multicolumn{3}{c}{Normal}  & \multicolumn{3}{c}{Anxious} \\ \hline
\multirow{3}{*}{MMPsy-Anxiety}  & Audio &\textbf{0.87} & 0.52 & 0.10 &\textbf{0.95}  & 0.91 & \textbf{0.95} & \textbf{0.93}
 & 0.16 & 0.10 & 0.12 \\ 
& Text &0.85 & 0.74 & 0.61 &0.88  & 0.96 & 0.88 & 0.92 & 0.34 & 0.61 & 0.43 \\ 
& Text+Audio & 0.85 & \textbf{0.76} & \textbf{0.65} &0.87  & \textbf{0.96} & 0.87 & 0.92 & \textbf{0.34} & \textbf{0.65} & \textbf{0.45} \\
\hline
\multicolumn{6}{c}{} & \multicolumn{3}{c}{Normal}  & \multicolumn{3}{c}{Depressed} \\ \hline
\multirow{3}{*}{MMPsy-Depression} & Audio &0.76 & 0.61 & 0.35 &0.86 & 0.84 & 0.86 & 0.85 & 0.39 & 0.35 & 0.37 \\ 
 & Text &0.81 & 0.74 & 0.62 &0.86  & 0.90 & 0.86 & 0.88 & 0.53 & 0.62 & 0.57 \\ 
& Text+Audio &\textbf{0.85} & \textbf{0.79} & \textbf{0.69} &\textbf{0.89}  & \textbf{0.92} & \textbf{0.89} & \textbf{0.9} & \textbf{0.61} & \textbf{0.69} & \textbf{0.64} \\
\hline
\end{tabular}
}
\caption{Ablation study on different modalities of our Mental-Perceiver on MMPsy. The best result is highlighted in \textbf{bold}.}
\label{modal}
\end{table*}

\subsection{Implementation Details}
We use Pytorch\footnote{http://pytorch.org} to implement our framework on Linux with two NVIDIA RTX 4090 GPU cards. The feature dimension $D_x$ is set to 768 and other dimensions $D_z$, $D_q$, and $D_y$ are all set to 512. In each epoch, all training data is shuffled randomly and then cut into mini-batches. The text feature and audio feature are concatenated as the multimodal input. We deploy the AdamW~\cite{loshchilov2017decoupled} optimizer for model optimization. We trained models for 200 epochs with an initial learning rate of 0.00003 and used LambdaLR to adjust the learning rate during training. The early stopping with patience 15 is deployed to accelerate training. We use the validation set for model selection and report the performance on the test set.

\subsection{Experiment Results}
\subsubsection{Main Results}
The experiment results of our Mental-Perceiver and various baselines on  MMPsy-Anxiety, MMPsy-Depression, and DAIC-WOZ are shown in Table~\ref{mainres}. From the results on different datasets, we can draw the following conclusions.  \textbf{1) On MMPsy-Anxiety}, our Mental-Perceiver outperforms baselines on Acc and UAR while achieving competitive performance on Sens and Spec. This shows that our Mental-Perceiver can achieve better overall anxiety detection performance with a better trade-off between Sensitivity and Specificity. For different categories, we can observe that our Mental-Perceiver achieves relatively high performance in the normal category and achieves the best performance in the anxious category. \textbf{2) On MMPsy-Depression}, our Mental-Perceiver outperforms baselines on Acc, UAR, and Sens while achieving competitive performance on Spec. This shows that our Mental-Perceiver can achieve better overall depression detection performance with a better trade-off between Sensitivity and Specificity. For different categories, we can observe that our Mental-Perceiver achieves the best precision and F1-score in the normal category while outperforming all baselines in the depressed category on all three metrics. \textbf{3) On DAIC-WOZ}, a similar conclusion can be reached. The Mental-Perceiver outperforms baselines on Acc, UAR, and Spec. Although the Mental-Perceiver's sensitivity is lower than the baseline ConvLSTM, ConvLSTM is very poor in specificity, indicating that there is a high rate of misdiagnosis on ConvLSTM. According to the UAR, we can observe Mental-Perceiver has a better balance between the false positive rate ( 1 - Specificity) and the false negative rate (1 - Sensitivity). Besides, according to the metrics on different categories, we can observe that the Mental-Perceiver achieves the best overall performance in both two categories. 

Overall, our Mental-Perceiver can achieve the best overall performance on different datasets for different mental disorder detection, showing the effectiveness and universality of our Mental-Perceiver for detecting different mental disorders. Besides, different models can achieve varying degrees of performance, indicating the effectiveness and usability of our MMPsy dataset as a benchmark for developing and evaluating mental disorder detection models.

\subsubsection{Ablation study on different modalities}
To verify the superiority of multimodal text-audio for mental disorder detection, we conduct an ablation study by using only audio, only text, and text+aduio as input to the Mental-Perceiver. The experimental results are shown in Table~\ref{modal}. We can observe that the Mental-Perceiver with multimodal inputs can achieve the best performance across various metrics on the MMPsy-Depression while achieving the best performance on UAR, Sensitivity, Precision in the normal category, Precision in the anxious category, Recall in the anxious category, and F1 on the anxious category and reasonable and competitive performance on Acc, Specificity, Recall on the normal category, and F1 on the normal category. Overall, multimodal input helps detect mental disorders.

\begin{table}[t]
\centering
\resizebox{1\linewidth}{!}{
\begin{tabular}{c|c|cccc}
\hline
Datasets & Modalities &Acc & UAR & Sens &Spec  \\ \hline
\multirow{3}{*}{MMPsy-Anxiety} & Mental-Perceiver & \textbf{0.85} & \textbf{0.76} & \textbf{0.65} & \textbf{0.87}  \\ 
&-Category Prior &0.84  &0.72 &0.58 & 0.86 \\
&-Matching Loss & 0.82 & 0.74 &0.62 & 0.86\\
\hline

\multirow{3}{*}{MMPsy-Depression} 
& Mental-Perceiver &\textbf{0.85} & \textbf{0.79} & \textbf{0.69} &\textbf{0.89} \\
&-Category Prior &0.82 &0.68 &0.49 &0.88 \\
&-Matching Loss &0.8 &0.72 &0.57 &0.86 \\
\hline
\end{tabular}
}
\caption{Ablation study on the effects of category priors and match loss on MMPsy. The best result is highlighted in \textbf{bold}.}
\label{loss}
\end{table}

\subsubsection{The effects of category priors and matching loss}
To investigate the effects of the category priors and matching loss, we conduct a study by removing the category priors and matching loss. The results are shown in Table~\ref{loss}. It can be seen that each component can improve the performance across various metrics, showing the effectiveness of category priors and matching loss.

\section{Conclusion}
In this work, we construct a new large-scale \textbf{M}ulti-\textbf{M}odal \textbf{Psy}chological assessment corpus (MMPsy) about anxiety and depression in adolescents who speak Mandarin. The MMPsy contains audios and extracted transcripts of responses from anxious/depressed and non-anxious/non-depressed adolescent volunteers. 
There are 7,736 cleaned interview data for anxiety detection and 4,247 for depression detection.
We further propose a novel mental disorder estimation network, named \textbf{Mental-Perceiver}, to detect anxious/depressive mental states automatically according to users' audio and corresponding transcripts. Extensive experiments on MMPsy and the public DAIC-WOZ show the effectiveness and superiority of our proposed Mental-Perceiver.

\section{Acknowledgements}
This work was supported in part by the National Natural Science Foundation of China (NSFC) under Grant No. 62206314, GuangDong Basic and Applied Basic Research Foundation under Grant No. 2022A1515011835, Science and Technology Projects in Guangzhou under Grant No. 2024A04J4388 and China Postdoctoral Science Foundation funded project under Grant No. 2021M703687.

\bibliography{aaai25}

\end{document}